\documentclass[revtex,12pt,pra,a4]{article}
\usepackage[dvipdfm]{graphicx}
\usepackage[usenames,dvipsnames]{xcolor}
\usepackage[dvipdfm]{hyperref}
\usepackage{amsmath}
\usepackage[makeroom]{cancel}
\usepackage{cite}
\begin{document}

\title{Quantized thermal current vortices}
\author{Ferenc M\'arkus$^{1,2}$\footnote{Corresponding author; email: markus.ferenc@ttk.bme.hu}  and Katalin Gamb\'ar$^{3,4}$\footnote{gambar.katalin@uni-obuda.hu} \\
$^{1}$Department of Physics, \\ Budapest University of Technology and Economics, \\ Műegyetem rkp. 3., H-1111 Budapest, Hungary \\ 
$^{2}$Department of Geography and Natural Sciences \\ Istv\'an Nemesk\"urty Faculty of Teacher Training \\  National University of Public Service, \\ Ludovika t\'er 2, H-1083 Budapest, Hungary  \\ 
$^{3}$Department of Natural Sciences, Institute of Electrophysics, \\ K\'alm\'an Kand\'o Faculty of Electrical Engineering,
\'Obuda University, \\ Tavaszmez\H{o} u. 17, H-1084 Budapest, Hungary \\ 
$^{4}$Department of Natural Sciences, National University of Public Service, \\ Ludovika t\'er 2, H-1083 Budapest, Hungary}

\maketitle

\begin{abstract}
The thermal Hall effect has emerged as a fundamental tool for probing exotic quasiparticles and topological order, particularly in magnetic insulators where electronic conduction is suppressed. Much like skyrmions, which are characterized by their topologically protected spin configurations, the thermal Hall effect is deeply rooted in the geometric properties of the underlying physical space.
Although the effect is a well-established experimental phenomenon, current research points toward the existence of its quantum analogue: the quantized thermal Hall effect. In this paper, we provide a theoretical framework for this quantum version based on Sommerfeld’s flux quantization. Furthermore, we demonstrate the potential existence of dissipationless thermal current vortices. We suggest that these vortices may play a crucial role in the stability and dynamics of other topological structures, such as skyrmion lattices, offering a new perspective on the interplay between heat transport and magnetic textures.
\end{abstract}

\section{Introduction}

In classical electromagnetism, the trajectory of a charged particle is deflected by a magnetic field through the Lorentz force. It is a compelling phenomenon that a thermal analogue of this effect exists, wherein heat flow undergoes a similar transverse deflection despite the absence of net electric charge transport. Unlike the Lorentz force-driven motion of electrons, the origin of this thermal Hall effect (or Righi--Leduc effect which can be created by phonons, magnons, or other exotic particles, e.g. Majorana) \cite{strohm2005,inyushkin2007,li2012} lies in more intricate mechanisms, such as the topological properties of the material and the Berry curvature of its excitations \cite{katsura2010,saito2019}. In a magnetic field, the Bloch phonon wavefunctions acquire a phase (Berry phase) that creates a curvature in $k$-space. This adds an anomalous velocity term to their motion, which deflects them perpendicular to the temperature gradient \cite{qin2012,zhang2010}.  Understanding this heat-current deviation on microscopic level is crucial for exploring emergent quasiparticles, particularly in systems where traditional electronic transport is suppressed. Recent studies show that the angular momentum of phonons (the chirality of phonons) and the coupling of the magnetic field may play an important role in the recognition of novel, phononmagnetic phenomena \cite{juraschek2025,ataei2024,lopez2026}.

The thermal Hall angle is the ratio of the transverse and longitudinal thermal conductivity $\theta_{\text{H}}=\kappa_{xy}/\kappa_{xx}$ \cite{strohm2005}, and its value, based on experimental results, typically ranges between $\theta_{\text{H}} \sim (10^{-4} -10^{-3}) T^{-1}$ in black phosphorus \cite{li2023};  $\theta_{\text{H}}$ is proportinal to the (weak) magnetic field.  In the measurements for rare-earth garnets (Tb$_3$Gd$_5$O$_{12}$) the obtained Hall angle is approximately  $5 \times 10^{-4} T^{-1}$ at 5 K temperature \cite{mori2014}. If we take the phonon free path ($l$) as the length scale, the radius can be estimated as follows: $R \approx l/\theta_{\text{H}}$. Since the phonon mean free path in pure crystals at low temperatures is around $10^{-6}$ m, the corresponding radius of curvature is on the order of 10 mm, estimated with a Hall angle of $10^{-4}$. This radius of curvature is macroscopic in size, as indicated by the measurement geometries. The semiclassical calculations also attempt to support the measured macroscopic results. However, the question arises as to whether a quantum-scale version of the phenomenon could exist? We will try to answer this with an interesting proposition. The sections of the article are organized as follows. In Sec. \ref{A classical frame of thermal Hall effect} we formulate the deflection effect in the analogy of the Lorentz force. With this, the formulated law takes on the usual form. In Sec. \ref{Properties of the deflected current}, we examine some properties of the deflected thermal current. The quantum property is introduced through the Sommerfeld magnetic flux quantization procedure in Sec. \ref{The role of the flux quantization}. The result, the appearance of quantized thermal current vortices, is also discussed as a conclusion. A possible connection point with regard to vorticity is the existence of Bloch-type magnetic skyrmions. It may happen that thermal vortices can make a significant contribution to the stability of the latter. This is described in Sec.  \ref{Discussion - Contribution to the skyrmions thermal stbility}. Lastly, we summarize our results in Sec. \ref{Summary}

\section{A classical framework of thermal the Hall effect}  \label{A classical frame of thermal Hall effect}

The model we propose is based on an analogy of a similar phenomenon that is already known. It is perhaps no coincidence that in the description of natural processes that are considered independent of each other, certain groups of phenomena are described by the same mathematical construction. We will now exploit this.

\begin{figure}[h]
\centering
\includegraphics[width=10 cm]{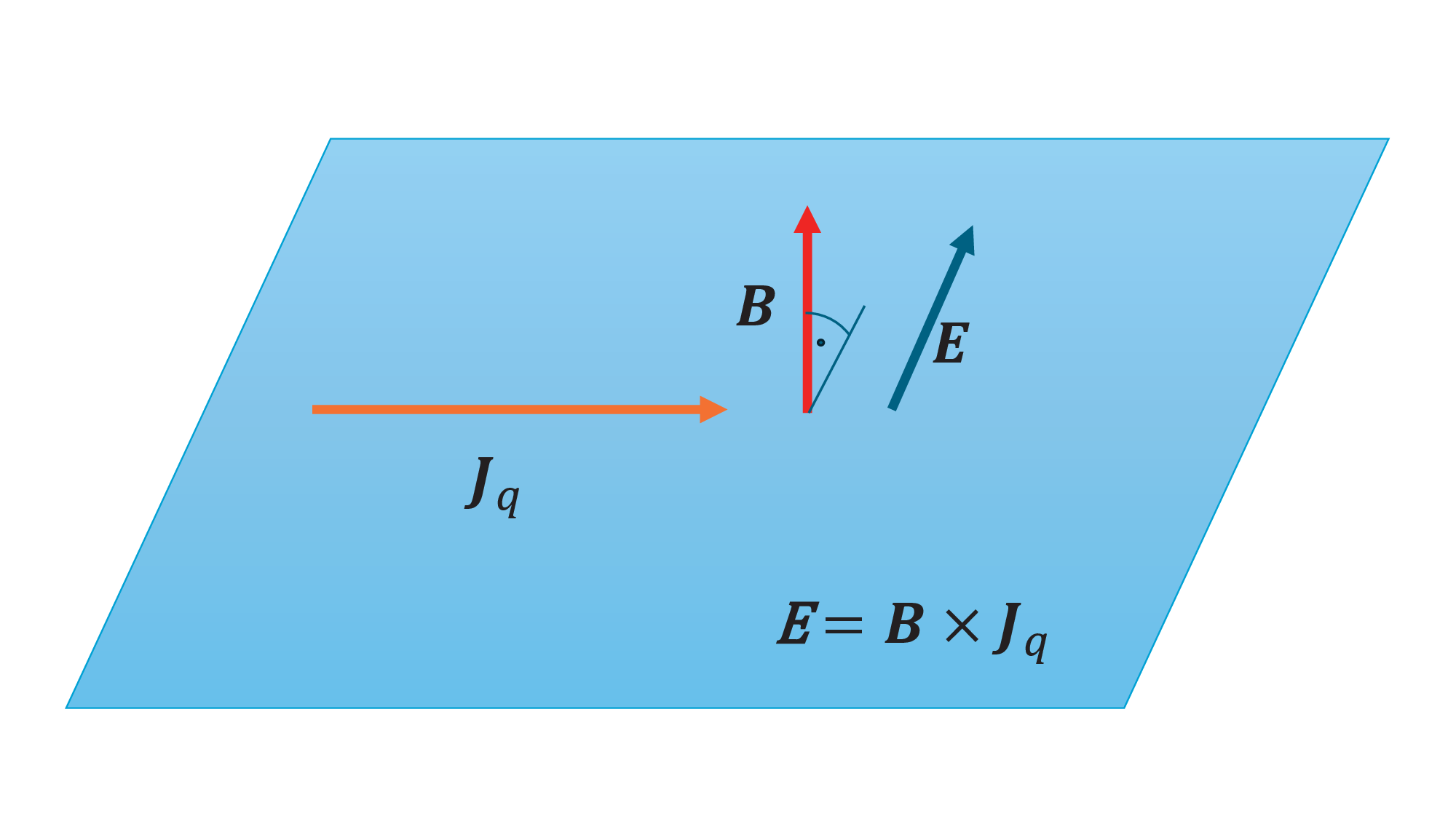}
\caption{The experimentally recognized and verified thermal Hall effect \cite{strohm2005,inyushkin2007,li2012}. Similarly to the electric Hall effect, the magnetic field ${\bf B}$ (in red; this can be external or it can come from the magnetization of the material itself) acts the heat current ${\bf J}_q$ (in orange). The measure of interaction is the exerting "action" is ${\bf E} $ (in green).}  
\label{thermal_Lorentz_1}
\end{figure}

To describe the thermal Hall effect a simple phenomenological picture can be formed if we restrict our comparison to the electric current Hall effect. We apply the analogy of the Lorentz formula, i.e., the vectorial product of electric current and magnetic field: ${\bf F} = q{\bf v} \times {\bf B}$. Let the heat flux ${\bf J}_q$ flow in the $x$-axis direction in the $(x,y)$ plane, with a uniform magnetic field ${\bf B}$ on the surface applied in the $z$-direction. Then the deflection effect on the heat flux, denoted by ${\bf E}$, is in the $y$-direction, as shown in Fig. \ref{thermal_Lorentz_1}.

We can mathematically formulate the phenomenon in the form 
\begin{equation}
{\bf E} = {\bf B} \times {\bf J}_q,   \label{thermal_Hall_force}
\end{equation}
which resembles the Lorentz force of moving charged particle in magnetic field. The relationship can be interpreted as an "action" ${\bf E}$ acting on the thermal current density ${\bf J}_q$ in a field of magnetic induction ${\bf B}$. The direction of the force is given by the cross product. The force ${\bf E}$ during the time $\Delta t$ changes the thermal current by $\Delta{\bf  J}_{\perp} $ in the direction of the force
\begin{equation}
{\bf E} \Delta t = \frac{1}{\varepsilon} \Delta {\bf J}_{\perp},  \label{thermal_Hall_pulse}
\end{equation}
where the $\varepsilon$ parameter is to match the units. The deviation of the heat flow rate follows from the two relationships
\begin{equation}
\frac{\Delta {\bf J}_{\perp}}{\Delta t} = \varepsilon  {\bf B} \times {\bf J}_q.  \label{thermal_Hall_effect}
\end{equation}

\begin{figure}[h]
\centering
\includegraphics[width=10 cm]{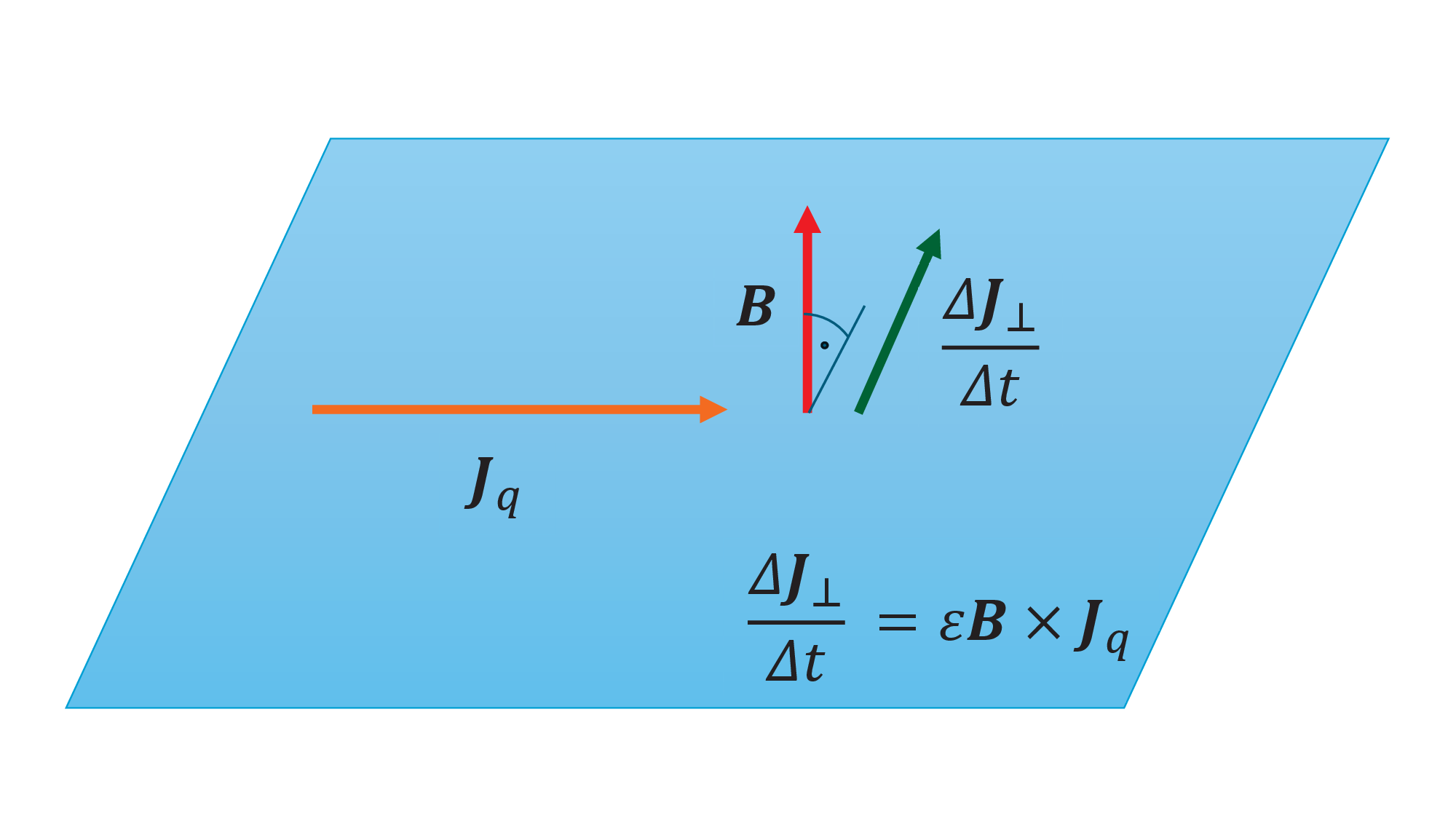}
\caption{In the thermal Hall effect, the change in the direction of the heat flow per unit time is marked in green. The parameter $\varepsilon$ characterizes the magnitude of the effect. }  
\label{thermal_Lorentz_2}
\end{figure}

This situation is shown in Fig. \ref{thermal_Lorentz_2}.

\section {Properties of the deflected current}  \label{Properties of the deflected current}

According to the experimental setups \cite{inyushkin2007} , when a temperature difference is applied in the $x$-direction, a heat flux, ${\bf J}_q$,  will flow parallel to it as a consequence. Mathematically, the thermal current is the gradient of temperature along the surface
\begin{equation}
{\bf J}_q = \kappa \nabla T  \label{heat_current},
\end{equation}
where $\kappa$ is the heat conductivity. The magnetic field ${\bf B}$ can be considered homogeneous on the surface
\begin{equation}
{\bf B} = (0,0,B_0),
\end{equation}
where $B_0$ means a constant field component in the $z$ axis. Let the components of the thermal current be
\begin{equation}
{\bf J}_q = (J_0,0,0),
\end{equation}
thus the thermal Hall current change per unit time is
\begin{equation}
\frac{\Delta {\bf J}_{\perp}}{\Delta t} \sim (0,\varepsilon B J_0,0).
\end{equation}
Due to the homogeneous magnetic field on the surface and the heat current given by the gradient of temperature in Eq. (\ref{heat_current}), the divergence of the (steady-state) perpendicular thermal current is 
\begin{equation}
\nabla \cdot {\bf J}_{\perp} \sim \nabla \cdot \left( {\bf B} \times {\bf J}_q \right) =  (\nabla \times {\bf B} ) \cdot {\bf J}_q - {\bf B} \cdot \nabla \times {\bf J}_q =0. \label{source_free_perp_thermal_current}
\end{equation}
Hence, ${\bf J}_{\perp}$ is a source-free thermal current. This highlights an important point that this source-free thermal flow vector space is also vortical such as the magnetic field.

\begin{figure}[h]
\centering
\includegraphics[width=10 cm]{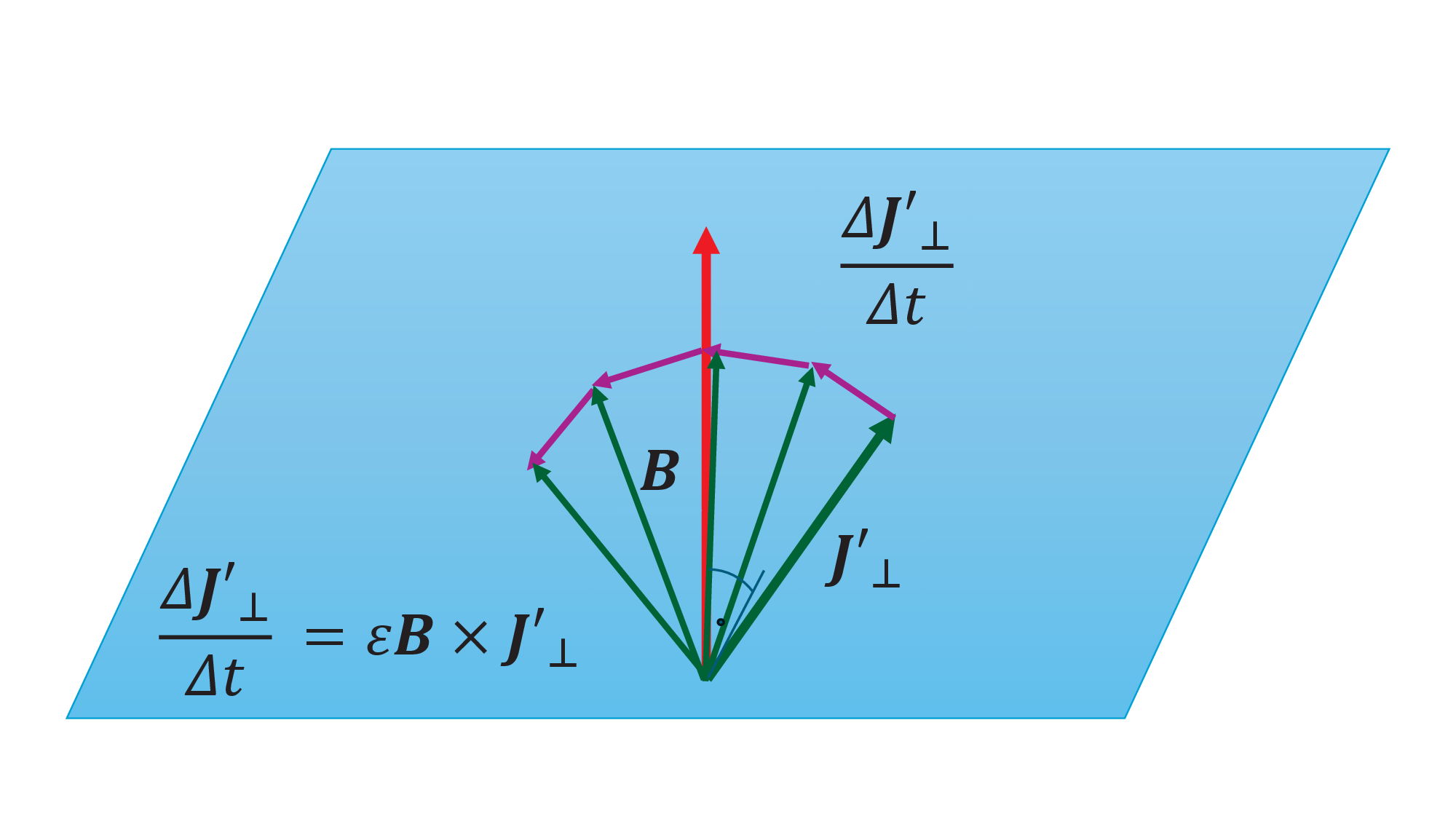}
\caption{The formation of a heat vortex is due to the thermal Hall effect. The heat flow is continuously deflected by the external magnetic field ${\bf B}$ (in red), resulting in a self-closing, non-dissipative heat flow. The thermal vortex is indicated by the purple color.}  
\label{thermal_Lorentz_3}
\end{figure}

We have just shown how the deflected current can be described. Now let us assume that the above relationship is also valid for the already deflected current ${\bf J}'_{\perp}$. We will apply a successive approximation for the thermal flow, i.e. The magnetic field also acts on the $ {\bf J}'_{\perp}$ current as in Eqs. (\ref{thermal_Hall_force}) -- (\ref{thermal_Hall_effect})

\begin{equation}
\frac{\Delta {\bf J}'_{\perp}}{\Delta t} =\varepsilon {\bf B} \times {\bf J}'_{\perp} .  \label{Hall_current_circle_1}
\end{equation}
The $\Delta {\bf J}'_{\perp}$ is a tangential thermal current, originating from ${\bf J}'_{\perp}$, and it flows on the cicumference of a circle, as shown in Fig. \ref{thermal_Lorentz_3}.

The thermal current in Eq. (\ref{Hall_current_circle_1}) also deviates under the influence of a magnetic field. If we consider a small $\Delta{\boldsymbol \varphi}$ rotation, the change in ${\bf J}'_{\perp}$ is
\begin{equation}
\Delta {\bf J}'_{\perp} = \Delta {\boldsymbol \varphi} \times {\bf J}'_{\perp}.
\end{equation}
After dividing by $\Delta t$ we get
\begin{equation}
\frac{\Delta {\bf J}'_{\perp}}{\Delta t} = {\boldsymbol \omega} \times {\bf J}'_{\perp}   \label{Delta_J_perp},
\end{equation}
where ${\boldsymbol \omega}$ is the angular frequency of the precession of ${\bf J}'_{\perp} $. This signs the appearance of a thermal vortex. This movement, however, suggests that it is not actually a thermal current driven by a temperature difference. Although the original driving force comes from a temperature gradient, the thermal Hall current is more like a heat motion, in which the related thermal energy density moves with velocity $v$, i.e., this is non-dissipative convective flow
\begin{equation}
{\bf J}'_{\perp} \sim {\bf v}.  \label{thermal_convective_current}
\end{equation}
Comparing this equation with Eq. (\ref{Hall_current_circle_1}) we can recognize a Larmor precession-like relation
\begin{equation}
{\boldsymbol \omega} = \gamma {\bf B},
\end{equation}
where $\gamma$ corresponds to the gyromagnetic ratio. Furthermore, this thermal flow also carries an entropy current
\begin{equation}
{\bf J}'_{s\perp} = \frac{{\bf J}'_{\perp}}{T},
\end{equation}
which is also a vortex current. Assuming a very small region of the surface to be of homogeneous temperature, based on Eq. (\ref{source_free_perp_thermal_current}) we can write 
\begin{equation}
\nabla \cdot {\bf J}'_{s\perp}  = 0,
\end{equation}
which means an entropy production-free, dissipationless vortex flow.

\section{The role of the flux quantization - thermal current vortices}  \label{The role of the flux quantization}

The magnetic field is generated by a vector potential, ${\bf A}$, acting in the plane of the surface as ${\bf B} = \nabla \times  {\bf A}$. The vector potential field is a vortex field. On the other hand, the effect that changes the direction of the vector ${\bf J}'_{\perp}$ to create its vortex current field can only be the vector potential ${\bf A}$ 
\begin{equation}
{\bf J}'_{\perp} \sim {\bf A}    \label{vector_potential}.
\end{equation}
Discrete values can be interpreted through the Sommerfeld magnetic flux quantization \cite{london1948,deaver1961}, which states that 
\begin{equation}
-e \oint {\bf A} {\textrm{d}} s=  n h.
\end{equation}
The semiclassical approximation works for Bloch functions and is therefore a viable option \cite{alexandradinata2018}. Comparing it with Eq. (\ref{vector_potential}) this  generates $n$ quantized thermal vortex currents
\begin{equation}  \label{quantized_thermal_current}
\oint {\bf J}'_{\perp} {\textrm{d}} s = -\eta \frac{h}{e} n,
\end{equation}
where we introduce the parameter $\eta$ to match the units. For the dissipation-free quantized entropy current vortices we can formulate
\begin{equation}
 \oint {\bf J}'_{s\perp} {\textrm{d}} s = -{\eta}{T} \frac{h}{e} n =  -{\eta}{T} \Phi_0 n,
\end{equation}
where the magnetic flux quantum is $\Phi_0= h/e$, since the quantization is associated with only one electron's charge. This is the main point of the quantized thermal Hall effect. After completing the integration in Eq. (\ref{quantized_thermal_current}), we obtain
\begin{equation}
(J'_{\perp} r)_n = - \frac{\eta \hbar}{e} n = - \frac{\eta \Phi_0}{2 \pi} n.  \label{quantized_J_times_r}
\end{equation}
Now the question arises, how do the factors in the product participate in the quantization process? A suitable description might be the application of Bohr-Landau quantization. The quantized angular momentum of the electron in a circular orbit is
\begin{equation}
m v r = n \hbar.
\end{equation}
On the other hand, we can write the Lorentz force holding the electron as
\begin{equation}
m \frac{v^2}{r} = evB.
\end{equation}
From these, the possible orbit radii are
\begin{equation}
r_n = \sqrt{\frac{\hbar}{eB}} \sqrt{n}.
\end{equation}
This means that the heat flow can only flow along the circumference of a circle with radii corresponding to the values of the quantum number $n$. The relevant Bohr radii are proportional to $\sqrt{n}$ thus we assume the same relation here
\begin{equation}
r_n =r_0 \sqrt{n}.
\end{equation}
Applying this in Eq. (\ref{quantized_J_times_r}) we obtain the expression for the quantized thermal currents
\begin{equation}
J'_{\perp n} = - \frac{ \eta \hbar }{e r_0}  {\sqrt{n}}.
\end{equation}
The above two relationships were derived from a kind of analogy. However, it is not the only possible choice. Here we can start from the fact that nature shows a kind of self-similarity in different disciplines.

The macroscopic measurement data for the thermal Hall radius show about 10 mm \cite{mori2014}. If we assume quantum properties, this radius value can change significantly. Li \emph{et al}. suggested a characteristic length \cite{li2023,sharma2024} on quantum level

\begin{equation}
\lambda_{tha} = \sqrt{\frac{h}{eB} \frac{\kappa_{xy}}{\kappa_{xx}}}
\end{equation}

The $\lambda_{tha}$ values calculated from measurement data for various insulators range between $0.2 - 0.7$ nm, regardless of the fact that the phonon mean free path in these materials varies by orders of magnitude. It is likely that this value greatly underestimates the true value. In any case, we can assume that the $1-10$ nm scale is realistic for thermal vortices to form. These phonons can be placed in a circular orbit with radius $r_0$ calculated above. This means that an estimate of the radius can be given based on the measurement results:

\begin{equation}
r_0 \sim  \sqrt{\frac{h}{eB} \frac{\kappa_{xy}}{\kappa_{xx}}}.
\end{equation}

Accordingly, the value of the resulting thermal current density is

\begin{equation}
J'_{\perp n} \sim - \eta \sqrt{\frac{hB}{e} \frac{\kappa_{xx}}{\kappa_{xy}}} {\sqrt{n}}.
\end{equation}

\begin{figure}[h]
\centering
\includegraphics[width=10 cm]{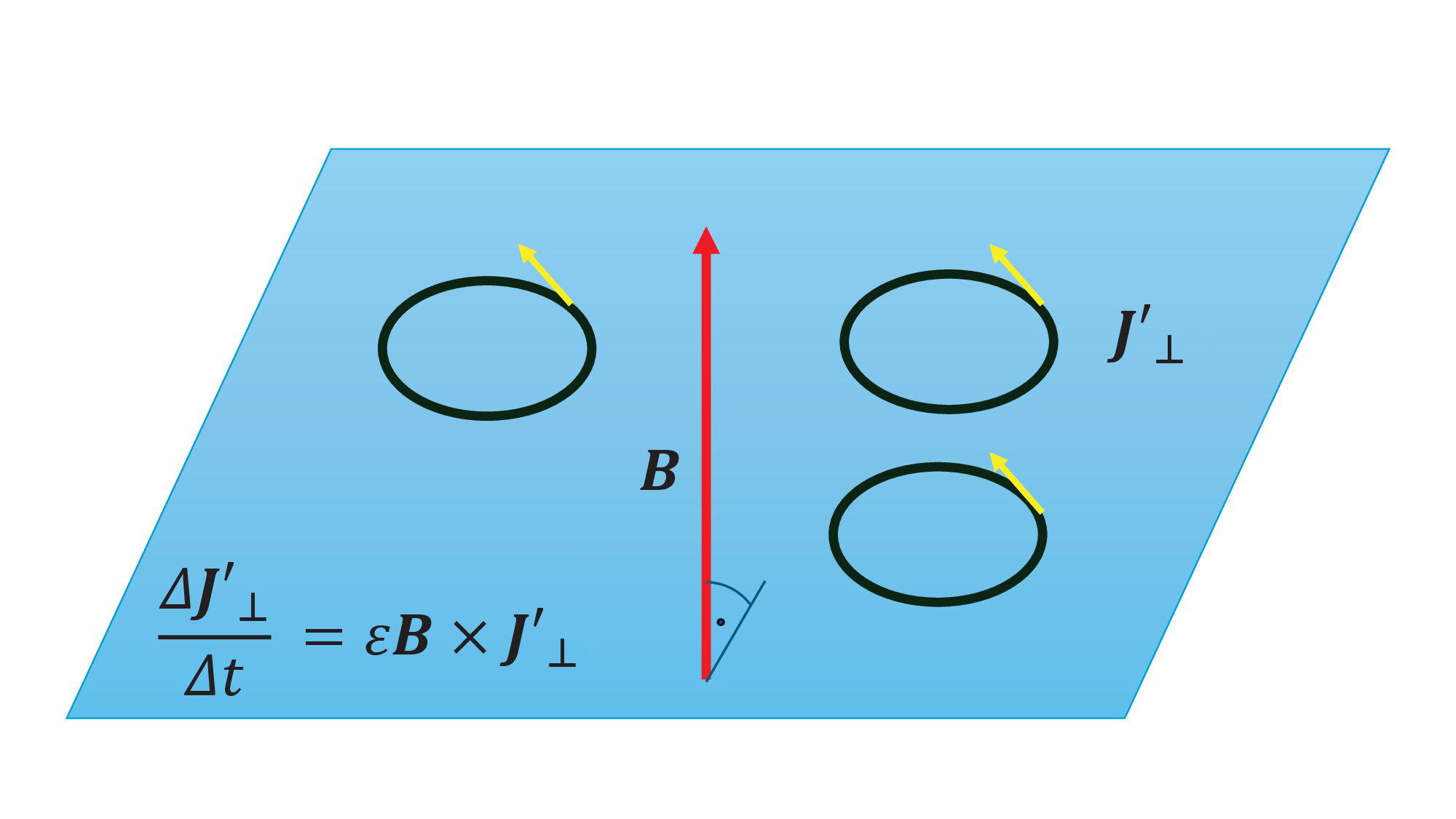}
\caption{Stable, non-dissipative thermal vortices (in green) formed in interaction with the magnetic field. The yellow arrows show the direction of the heat current.}  
\label{thermal_Lorentz_4}
\end{figure}

The magnitude of the current density depends only on the now unknown parameter $\eta$. We can conclude from the above considerations that these quantized convective thermal currents can be thought of as tiny vortices of ${\bf J}'_{\perp}$ that behave like particles in matter as shown in Fig. \ref{thermal_Lorentz_4} . This quantized appearance is somewhat similar to the topological protection of skyrmions \cite{bogdanov1989,muhlbauer2009}, which gives them extraordinary stability. Based on this, we can term them thermions. Their “knot-like” arrangement means they don’t break apart easily, a stable configuration that could make them ideal for future energy storage solutions.

\section{Discussion - Contribution to the skyrmions thermal stability}   \label{Discussion - Contribution to the skyrmions thermal stbility}

Since the thermal Hall effect is a topological phenomenon, the possibility of coupling with other, similarly manifested phenomena justifiably arises. A realistic connection possibility is based on the existence of vorticity. The thermal vortices raised can be related to other phenomena for structural reasons. Let us consider a seemingly completely independent set of phenomena. Bloch-type magnetic skyrmions are nanoscale, vortex-like spin textures characterized by an azimuthal (tangential) rotation of magnetization from core to periphery, typically found in B20-type chiral magnets \cite{muhlbauer2011}. These robust, topologically protected spin structures are driven by the Dzyaloshinskii-Moriya interaction (DMI) and can form, move, and rearrange in lattices under applied magnetic fields in the non-negligible thermal effects. Magnetic skyrmion domain structures are more complex than previously thought due to the new degree of freedom called helicity. Key findings indicate the discovery of helicity-reversals within magnetic walls and skyrmions, resulting in a unique multi-ring internal structure, while maintaining topological stability \cite{yu2012}. In a newer exprerimental  examination Birch \emph{et al}. \cite{birch2022} documented the breakdown of stripe domain structures into individual skyrmions in Fe$_3$GeTe$_2$ (FGT) layers under the influence of an external magnetic field, and change of the temperature. The typical size of skyrmions is approximately 10-100 nm, although in certain layered magnetic structures these can be as small as 1–5 nm \cite{jiang2017}. As we saw earlier, the estimated size range of non-dissipative thermal vortices is also in this range. This is important because in the present case the thermal effect does not try to break up the existing structure, but on the contrary, it may even stabilize it \cite{chalus2025}. This may arise from the fact that due to the local thermal vortical flow there is no equalization process similar to the macroscopic cases. It would certainly be interesting to investigate this experimentally. The case of nanotubes may be worth considering, since all physical quantities are quantized perpendicular to the axis, so presumably only \emph{thermions} with a radius that matches the radius of the nanotube can be produced. In addition, nanotubes exhibit interesting thermal behavior anyway \cite{hone1999,hone2002,kim2001,markusbg2018}. Similarly interesting is the investigation of the quantized thermal Hall effect realized in MXene Scrolls \cite{zhang2026}.

\section{Summary}  \label{Summary}

In analogy to the Hall effect describing the deflection of an electric current in a magnetic field, there is a similar phenomenon for heat flow. In the thermal Hall effect, the heat flow is deflected by the magnetic field, which has been experimentally confirmed. In this article, we show that this thermal Hall effect can create closed non-dissipative heat loops, or in other words, thermal vortices, \emph{thermions}. We also show that the quantized existence of this phenomenon is also possible and explainable.  The origin the effect based on the Sommerfeld quantization of the magnetic flux and the Bohr-Landau quantization of the angluar momentum. These thermal vortices only allow heat flows and orbital radii of the quantum number. Such closed vortices are also created by Bloch skyrmions in magnetic materials. In these cases, thermal effects usuallylead to instability. The goals of the future research direction include the possibility that thermal vortices actually stabilize skyrmions much better. If this is the case, then with the appropriate parameter settings a very positive effect can be achieved. This requires further investigation. 

The study highlights that the stability of skyrmions is significantly dependent on the magnetic history (protocol) used to reach a given state. Based on the thermal vortices we have discussed, it is suggested that this phenomenon may also contribute to achieving a stable skyrmion structure. In fact, here the thermal vortex effect does not destroy the formed skyrmion, but rather stabilize it. This can be determined by further experiments. \\

{\bf Acknowledgment}

TKP2021-NVA-16 has been implemented with the support provided by the Ministry of Innovation and Technology of Hungary from the National Research, Development and Innovation Fund. F. M. would like to thank NKKP 149457.


\begin{thebibliography}{100}

\bibitem{strohm2005} C. Strohm, G. L. J. A. Rikken, and P. Wyder, {\it Phenomenological Evidence for the Phonon Hall Effect}. Phys. Rev. Lett. {\bf 95}, 155901 (2005).
\bibitem{inyushkin2007} A. V. Inyushkin, and A. N. Taldenkov, {\it On the Phonon Hall Effect in a Paramagnetic Dielectric}. JETP Letters {\bf 86}, 379 (2007).
\bibitem{li2012} N. Li, J. Ren, L. Wang, G. Zhang, P. H\"anggi, and B. Li, {\it Colloquium: Phononics: Manipulating heat flow with electronic analogs and beyond}. Rev. Mod. Phys. {\bf 84}, 1045 (2012).

\bibitem{katsura2010} H. Katsura, N. Nagaosa, and P. A. Lee, {\it Theory of the Thermal Hall Effect in Quantum Magnets}. Phys. Rev. Lett. {\bf 104}, 066403 (2010).

\bibitem{saito2019} T. Saito, K. Misaki, H. Ishizuka, and N. Nagaosa, {\it Berry Phase of Phonons and Thermal Hall Effect in Nonmagnetic Insulators}. Phys. Rev. Lett. {\bf 123}, 255901 (2019).

\bibitem{qin2012} T. Qin, J. Zhou, and J. Shi, {\it Berry Curvature and the Phonon Hall Effect}. Phys. Rev. B{\bf 86}, 104305 (2012). 
\bibitem{zhang2010} L. Zhang, J. Ren, J.-S. Wang, and B. Li, {\it Topological Nature of the Phonon Hall Effect} Phys. Rev. Lett. {\bf 105}, 225901 (2010).

\bibitem{juraschek2025} D. M. Juraschek , R. M. Geilhufe, H. Zhu, M. Basini, P. Baum, A. Baydin, S. Chaudhary, M. Fechner, B. Flebus, G. Grissonnanche, A. I. Kirilyuk, M. Lemeshko, S. F. Maehrlein, M. Mignolet, S. Murakami, Q. Niu, U. Nowak, C. P. Romao, H. Rostami, T. Satoh, N. A. Spaldin, H. Ueda, and L. Zhang, {\it Chiral phonons}. Nature Physics {\bf 21}, 1532 (2025).

\bibitem{ataei2024} A. Ataei, G. Grissonnanche, M.-E. Boulanger, L. Chen, É. Lefrançois, V. Brouet, and L. Taillefer, {\it Phonon chirality from impurity scattering in the antiferromagnetic phase of Sr$_2$IrO$_4$}. Nature Physics {\bf 20}, 585 (2024).

\bibitem{lopez2026} D. A. B. Lopez,V. Brehm, and D. M. Juraschek, {\it Atomistic theory of the phonon angular momentum Hall effect}. arXiv (2026). 

\bibitem{li2023} X. Li, Y. Machida, A. Subedi, Z. Zhu, L. Li1, and K. Behnia, {\it The Phonon Thermal Hall Angle in Black Phosphorus}. Nature Comm. {\bf 14}, 1027 (2023).
\bibitem{mori2014} M. Mori, A. Spencer-Smith, O. P. Sushkov, and S. Maekawa, {\it Origin of the Phonon Hall Effect in Rare-Earth Garnets}. Phys. Rev. Lett. {\bf 113}, 265901 (2014).

\bibitem{london1948} F. London, {\it On the Problem of the Molecular Theory of Superconductivity}. Phys. Rev. {\bf 74}, 562 (1948).
\bibitem{deaver1961} B. S. Deaver, Jr. and W. M. Fairbank, {\it Experimental Evidence for Quantized Flux in Superconducting Cylinders}. Phys. Rev. Lett. {\bf 7}, 43 (1961). 

\bibitem{alexandradinata2018} A. Alexandradinata and Leonid Glazman, {\it Semiclassical Theory of Landau Levels and Magnetic Breakdown in Topological Metals} Phys. Rev. B {\bf 97}, 144422 (2018).

\bibitem{sharma2024}  R. Sharma, M. Valldor, and T. Lorenz, {\it Phonon thermal Hall effect in nonmagnetic Y$_2$Ti$_2$O$_7$}. Phys. Rev. B {\bf 110}, L100301 (2024).

\bibitem{bogdanov1989} A. N. Bogdanov and D. A. Yablonskii, {\it Thermodynamically stable "vortices" in magnetically ordered crystals. The mixed state of magnets.} Soviet Physics JETP {\bf 68}, 101 (1989).
\bibitem{muhlbauer2009} S. M\"uhlbauer, B. Binz, F. Jonietz, C. Pfleiderer, A. Rosch, A. Neubauer, R. Georgii, and P. B\"oni, {\it Skyrmion Lattice in a Chiral Magnet}. Science, {\bf 323}, 915 (2009).

\bibitem{muhlbauer2011} S. Mühlbauer, B. Binz, F. Jonietz, C. Pfleiderer,. Rosch, A. Neubauer, R. Georgii, P. Böni, {\it Skyrmion Lattice in a Chiral Magnet}. Science {\bf 323}, 915 (2011).

\bibitem{yu2012} X. Yu, M. Mostovoy, Y. Tokunaga, W. Zhang, K. Kimoto, Y. Matsui, Y. Kaneko, N. Nagaosa, and Y. Tokura, {\it Magnetic Stripes and Skyrmions with Helicity Reversals}. PNAS {\bf 109}, 8856 (2012).

\bibitem{birch2022} M. T. Birch, L. Powalla, S. Wintz, O. Hovorka, K. Litzius, J. C. Loudon, L. A. Turnbull, V. Nehruji, K. Son, C. Bubeck, T. G. Rauch, M. Weigand, E. Goering, M. Burghard, and G. Schütz, {\it History-dependent domain and skyrmion formation in 2D van der Waals magnet Fe$_3$GeTe$_2$}. Nat. Com. {\bf 13}, 3035 (2022).
\bibitem{jiang2017} W. Jiang, G. Chen, K. Liu, J. Zang, S. G.E. te Velthuis, and Axel Hoffmann, {\it Skyrmions in Magnetic Multilayers}. Phys. Rep. {\bf 704}, 1 (2017) .

\bibitem{chalus2025} N. Chalus, A. W. D. Leishman, R. M. Menezes, G. Longbons, U. Welp,4 W.-K. Kwok, J. S. White, M. Bartkowiak, R. Cubitt, Y. Liu, E. D. Bauer, M. Janoschek, M. V. Milo\u{s}evi\'c, and M. R. Eskildsen, {\it Skyrmion Lattice Manipulation with Electric Currents and Thermal Gradients in MnSi}. Phys. Rev. B {\bf 111}, 064410 (2025).

\bibitem{hone1999} J. Hone, M. Whitney, C. Piskoti, and A. Zettl, {\it Thermal Conductivity of Single-Walled Carbon Nanotubes}. {Phys. Rev. B} {\bf 59}, R2514, (1999).

\bibitem{hone2002} J. Hone, M.C. Llaguno, M.J. Biercuk, A.T. Johnson, B. Batlogg, Z. Benes, and J. E. Fischer, {\it Thermal properties of carbon nanotubes and nanotube-based materials}. Appl. Phys. A {\bf 74}, 339, (2002).

\bibitem{kim2001} P. Kim, L. Shi, A. Majumdar, and P. L. McEuen, {\it Thermal Transport Measurements of Individual Multiwalled Nanotubes}. {Phys. Rev. Lett.} {\bf 87}, 215502, (2001).

\bibitem{markusbg2018} B. G. M\'arkus, B. Gy\"ure-Garami, O. S\'agi, G. Cs\H{o}sz, A. Karsa, F. Márkus, and F. Simon, {\it Heating Causes Nonlinear Microwave Absorption Anomaly in Single-Walled Carbon Nanotubes}, Physica Status Solidi {\bf 255}, 1800258 (2018).

\bibitem{zhang2026} T. Zhang, B. Chacon, D. Zhang, A. Cotton, Y. Zhang, Y. Zhang, S. Ippolito, F. Urban, T. Parker, L. Bi, K. Shevchuk, K. Matthews, E. A. Stach, and Y. Gogotsi, {\it Scalable Synthesis of MXene Scrolls}. Adv. Materials {\bf 38}, 21457 (2026).

\end{thebibliography}
\end{document}